\documentclass[onecolumn,showpacs]{revtex4}

\usepackage{graphicx}
\usepackage{dcolumn}
\usepackage{bm}

\input epsf

\begin{document}

\title{\Large Holographic Dark Energy Interacting with Two Fluids and Validity of Generalized Second Law of Thermodynamics}

\author{\bf Ujjal Debnath$^1$\footnote{ujjaldebnath@yahoo.com ,
ujjal@iucaa.ernet.in}}

\affiliation{$^1$Department of Mathematics, Bengal Engineering and
Science University, Shibpur, Howrah-711 103, India.}

\date{\today}

\begin{abstract}
We have considered a cosmological model of holographic dark energy interacting with
dark matter and another unknown component of dark energy of the universe. We have
assumed two interaction terms $Q$ and $Q'$ in order to include the scenario in which the
mutual interaction between the two principal components (i.e., holographic dark
energy and dark matter) of the universe leads to some loss in other forms of cosmic
constituents. Our model is valid for any sign of $Q$ and $Q'$. If $Q<Q'$, then part
of the dark energy density decays into dark matter and the rest in the other unknown
energy density component. But if $Q>Q'$, then dark matter energy receives from dark
energy and from the unknown component of dark energy. Observation suggests that
dark energy decays into dark matter. Here we have presented a general prescription of a
cosmological model of dark energy which imposes mutual interaction between holographic
dark energy, dark matter and another fluid. We have obtained the equation of state for
the holographic dark energy density which is interacting with dark matter and other
unknown component of dark energy. Using first law of thermodynamics, we have obtained the
entropies for holographic dark energy, dark matter and other component of dark energy, when
holographic dark energy interacting with two fluids (i.e., dark matter and other component of
dark energy). Also we have found the entropy at the horizon when the radius ($L$) of the event
horizon measured on the sphere of the horizon. We have investigated the GSL of thermodynamics
at the present time for the universe enveloped by this horizon. Finally, it has been obtained
validity of GSL which implies some bounds on deceleration parameter $q$.

\end{abstract}


\maketitle

\section{\normalsize\bf{Introduction}}

Recent observation of the luminosity of type Ia supernovae
indicate [1, 2] an accelerated expansion of the universe and the
surveys of clusters of galaxies show that the density of matter is
very much less than the critical density. This observation leads
to a new type of matter which violate the strong energy condition
$\rho+3p<0$. The matter content responsible for such a condition
to be satisfied at a certain stage of evaluation of the universe
is referred to as dark energy [3-6]. This mysterious fluid is
believed to dominate over the matter content of the Universe by 70
$\%$ and to have enough negative pressure as to drive present day
acceleration. Most of the dark energy models involve one or more
scalar fields with various actions and with or without a scalar
field potential [7]. On the other hand when the universe was
380,000 years old neutrinos was 10{\%} atoms i.e. usual baryonic
matter was 12{\%}, dark matter was 63{\%}, photons 15{\%} and dark
energy was negligible. In the analysis of dark energy the main
attraction should be on the state parameter $w=\frac{p}{\rho}$
where $p$ and $\rho$ are the pressure and energy density of the
dark energy. In Cosmological constant model $w=-1$ around present
epoch [8] from $w>-1$ in the near past [9]. There are various
kinds of models of dark energy and among all of them, the simplest
case is the $\Lambda CDM$ model which has $\rho = \Lambda$ =
constant and the equation of state as $p = - \rho$. It fits our
observational data very well but a problem called Coincidence
problem [10] arises in this model, which requires extreme
fine-tuning of the order of magnitude $10^{120}$ of the
initial value of $\Lambda$.\\

Now, as the observational data permits us to have a rather time
varying equation of state, there are a bunch of models
characterized by different scalar fields such as a slowly rolling
scalar field (Quintessence) [11], kinetic energy induced K-essence
[12], a tachyonic field [13], Chaplygin gas [14], a phantom model [15] and also a
quintom model [9]. In a phantom model, we have the equation of
state as $p  =  \omega  \rho$, where $\omega < - 1$. The simplest
type of phantom model is a scalar field having a potential
V($\phi$), the kinetic energy of which is negative [16]. The
quintom model has two scalar fields, one is like that of the
quintessence model and the other is like that of the phantom
model. The condition $\omega = - 1$ is named as the phantom
divide. There are even models which can smoothly cross this
phantom divide [17]. The speciality of a phantom model lies in the
fact that in these type of models the universe ends with a Big Rip
singularity [18], which means that in a finite time, $|p|
\rightarrow \infty$, $\rho \rightarrow \infty$ and $a \rightarrow
\infty$, where $a(t)$ is the scale factor. Also, the current
observation data from Type-Ia supernovae and the CMB anisotropy
documents give us limits to the various parameters [19-23] like
$\Omega_{B}$,  $\Omega_{DE}$, $\Omega_{DM}$ where  $\Omega$
denotes the relative density and the suffices $B,~ DE,~ DM$
represent baryonic matter, dark energy and dark matter
respectively. It also gives us data from which we have the limit $
- 1.38 < \omega < - 0.82 $ [24] with a very high level of
confidence where $\omega$ is the equation of state parameter.
Recent observations also reveals
the fact that our universe is likely to be spatially flat [25].\\

The holographic principle emerged in the context of black-holes,
where it was noted that a local quantum field theory can not fully
describe the black holes [26]. Some long standing debates regarding
the time evolution of a system, where a black hole forms and then
evaporates, played the key role in the development of the
holographic principle [27-29]. Cosmological versions of
holographic principle have been discussed in various literatures
[30-32]. Easther et al [32] proposed that the holographic
principle be replaced by the generalized second law of
thermodynamics when applied to time-dependent backgrounds and
found that the proposition agreed with the cosmological
holographic principle proposed by Fischler and Susskind (Ref [30])
for an isotropic open and flat universe with a fixed equation of
state. Verlinde [33] studied the holographic principle in the
context of an ($n+1$) dimensional radiation dominated closed FRW
universe. Numerous cosmological observations have established the
accelerated expansion of the universe [34,35]. Since it has been
proven that the expansion of the universe is accelerated, the
physicists and astronomers started considering the dark energy
cosmological observations indicated that at about 2/3 of the total
energy of the universe is attributed by dark energy and 1/3 is due
to dark matter [36]. In recent times, considerable interest has
been stimulated in explaining the observed dark energy by the
holographic dark energy model [26,36,37]. An approach to the
problem of dark energy arises from the holographic principle
stated in the first paragraph. For an effective field theory in a
box size $L$ with UV cutoff $\Lambda_{c}$, the entropy
$L^{3}\Lambda_{c}^{3}$. The non-extensive scaling postulated by
Bekenstein suggested that quantum theory breaks down in large
volume [36]. To reconcile this breakdown, Chohen et al [38]
pointed out that in quantum field theory a short distance (UV)
cut-off is related to a long distance (IR) cut-off due to the
limit set by forming a black hole. Taking the whole universe into
account the largest IR cut-off $L$ is chosen by saturating the
inequality so that we get the holographic dark energy density as
[36] $\rho_{\Lambda}=3c^{2}M_{p}^{2}L^{-2}$ where $c$ is a
numerical constant and $M_{p}\equiv1/\sqrt{8\pi G}$ is the reduced
Plank mass. On the basis of the holographic principle proposed by
[30] several others have studied holographic model for dark energy
[35]. Employment of Friedman equation [39] $\rho=3M_{p}^{2}H^{2}$
where $\rho$ is the total energy density and taking $L=H^{-1}$ one
can find $\rho_{m}=3(1-c^{2})M_{p}^{2}H^{2}$ for flat universe. Thus either
$\rho_{m}$ or $\rho_{\Lambda}$ behaves like $H^{2}$. Thus, dark
energy results as pressureless, since $\rho_{\Lambda}$ scales like
matter energy density $\rho_{m}$ with the scale factor $a$ as $a^{-3}$.
But, neither dark energy, nor dark matter has laboratory evidence for
its existence directly. Also, taking the apparent horizon as the IR
cut-off may result in a constant parameter of state $w$, which is in
contradiction with recent observations implying variable $w$ [40].
For small value of $\Omega_{k}$ in non-flat universe, Setare et al [41] have
considered a model as a system which departs
slightly from flat space. Consequently, the results for
the flat universe they treat the apparent horizon
only as an arbitrary distance and not as the system's IR cut-off.\\

Interaction models where the dark energy weakly interacts with
the dark matter have also been studied to explain the evolution
of the Universe. This models describe an energy flow between the
components. To obtain a suitable evolution of the Universe an interaction is
often assumed such that the decay rate should be proportional to
the present value of the Hubble parameter for good fit to the
expansion history of the Universe as determined by the Supernovae
and CMB data [42]. These kind of models describe an energy flow
between the components so that no components are conserved
separately. A variety of interacting holographic dark energy models have been
proposed and studied for this purpose [42-45].\\

Since the discovery of black hole thermodynamics in 1970,
physicists have speculated on the thermodynamics of cosmological
models in an accelerated expanding universe. In 1973, Bekenstein
[46] assumed that there is a relation between the event of horizon
and the thermodynamics of a black hole, so that the event of
horizon of the black hole is a measure of the entropy of it. This
idea has been generalized to horizons of cosmological models, so
that each horizon corresponds to an entropy. Thus the second law
of thermodynamics was modified in the way that in generalized
form, the sum of all time derivative of entropies related to
horizons plus time derivative of normal entropy must be positive,
i.e. the sum of entropies must be increasing function of time.
There is a cosmological event horizon, analogous to a black hole
horizon, which can be associated with thermodynamical variables.
Supposing that some energy passes through the cosmological event
horizon, the definitions of Black Hole temperature and entropy
imply that the first law of thermodynamics is valid. In the
semiclassical quantum description of black hole physics, it was
found that black holes emit Hawking radiation with a temperature
proportional to their surface gravity at the event horizon and
they have an entropy which is one quarter of the area of the event
horizon in Planck units [47]. The temperature, entropy and mass of
black holes satisfy the first law of thermodynamics [48]. On the
other hand, it was shown that the Einstein equation can be derived
from the first law of thermodynamics by assuming proportionality
of the entropy and the horizon area [49]. The Einstein equation
for the nonlinear gravitational theory $f(R)$ was also derived
from the first law of thermodynamics with some non-equilibrium
corrections [50]. For a general static spherically symmetric
space–time, Padmanabhan [51] showed that the Einstein equation at
the horizon gives the first law of thermodynamics on the horizon.
The study of the relation between the Einstein equation and the
first law of thermodynamics has been generalized to the
cosmological context where it was shown that the first law of
thermodynamics on the apparent horizon $\tilde{r}_{A}$ can be
derived from the Friedmann equation and vice versa if we take the
Hawking temperature and the entropy on the apparent horizon [52].\\

The thermodynamics in de Sitter space–time was first investigated by Gibbons and Hawking in [53]. In a
spatially flat de Sitter space–time, the event horizon and the apparent horizon of the
Universe coincide and there is only one cosmological horizon. When the apparent horizon
and the event horizon of the Universe are different, it was found that the first law and
the second law of thermodynamics hold on the apparent horizon, while they break down if one
considers the event horizon [54]. Thermodynamics of the expanding universe has
also been the subject of several studies [55-64]. Phantom thermodynamics looks
leading to negative entropy of the universe [65] or to appearance of negative temperatures
[66]. In accelerated expanding universe, besides the normal entropy,
a cosmological horizon entropy can also be considered. One can investigate the
conditions for which the generalized second law of thermodynamics (GSL) holds
[58, 59]. In these cases GSL asserts that the sum of the horizon entropy, and
the normal entropy of the fluid is an increasing function of time. In [58] the
change in event-horizon area in cosmological models that depart slightly from
de Sitter space was investigated, and it was shown that the area and consequently
the (de Sitter) horizon entropy are non decreasing functions of time. In
the presence of a viscous fluid, there was found that GSL was satisfied provided
that the temperature of the fluid was equal to or lower than de Sitter horizon
temperature. Gong et al [67] derived the temperature and entropy
of the matter contents inside the apparent horizon from the first law of thermodynamics
and discuss the holographic entropy bound and the generalized second law (GSL) of
thermodynamics for the Universe with DE. They have addressed the thermodynamics of DE by
considering the DE models with constant $w$ and the generalized Chaplygin gas (GCG).\\

In the present work, we have considered a cosmological model of holographic dark energy interacting with
dark matter and another unknown component of dark energy of the universe. We have
assumed two interaction terms $Q$ and $Q'$ in order to include the scenario in which the
mutual interaction between the two principal components (i.e., holographic dark
energy and dark matter) of the universe leads to some loss in other forms of cosmic
constituents. In section II, we have presented a general prescription of a
cosmological model of dark energy which imposes mutual interaction between holographic
dark energy, dark matter and another fluid. We have obtained the equation of state for
the holographic dark energy density which is interacting with dark matter and other
unknown component of dark energy. In section III, we have obtained the
entropies for holographic dark energy, dark matter and other component of dark energy, when
holographic dark energy interacting with two fluids (i.e., dark matter and other component of
dark energy). We have investigated the validity GSL of thermodynamics at the present time for
the universe enveloped by the horizon. Finally, we have presented some concluding remarks in section IV.\\

\section{\normalsize\bf{Holographic Dark Energy Interacting with Two Fluids}}

Assuming the universe to be homogeneous and isotropic, the
Friedmann-Robertson-Walker (FRW) metric can be written as

\begin{equation}
ds^{2}= dt^{2}-a^{2}(t)\left[\frac{dr^{2}}{1-kr^{2}}+
r^{2}(d\theta^{2}+\sin^{2}\theta  d\phi^{2})\right]
\end{equation}

where $a(t)$ is the expansion scalar or the scale factor and $k
~(= 0,\pm 1)$ is the curvature scalar. Then Einstein's field
equations become (choosing $8\pi G=c=1$)

\begin{equation}
3H^{2}+\frac{3k}{a^{2}}=\rho_{\Lambda}+\rho_{m}+\rho_{X}
\end{equation}
and
\begin{equation}
2\dot{H}-\frac{2k}{a^{2}}=-[(\rho_{\Lambda}+\rho_{m}+\rho_{X})+(p_{\Lambda}+p_{m}+p_{X})]
\end{equation}

where $\rho_{\Lambda},~\rho_{m},~\rho_{X}$ and
$p_{\Lambda},~p_{m},~p_{X}$ are respectively energy density and
pressure of holographic dark energy, dark matter and another unknown
component of dark energy. We will assume that the dark matter component
is interacting with the holographic dark energy component, so their continuity
equations take the form [68]

\begin{equation}
\dot{\rho}_{\Lambda}+3H(\rho_{\Lambda}+p_{\Lambda})=-Q'
\end{equation}
and
\begin{equation}
\dot{\rho}_{m}+3H(\rho_{m}+p_{m})=Q
\end{equation}

where $Q$ and $Q'$ in order to include the scenario in which the
mutual interaction between the two principal components of the
universe leads to some loss in other forms of cosmic constituents.
In this case, we have assumed $Q\ne Q'$, so the continuity
equation for other component of dark energy becomes

\begin{equation}
\dot{\rho}_{X}+3H(\rho_{X}+p_{X})=Q'-Q
\end{equation}

If $Q<Q'$, then part of the dark energy density decays into dark matter
and the rest in the other unknown energy density component. But if $Q>Q'$,
then dark matter energy receives from dark energy and from the unknown
component of dark energy.\\

We are taking about in this case that dark energy decay into dark
matter (or vice versa, depending on the sign of $Q$) and other
component. Assume, the interaction
terms $Q$ and $Q'$ are [69]

\begin{equation}
Q=\Gamma_{m}\rho_{\Lambda},~~Q'=\Gamma_{\Lambda}\rho_{\Lambda}
\end{equation}

where, $\Gamma_{\Lambda}$ is the decaying rate of energy from holographic dark energy to
dark matter and other unknown component of dark energy and $\Gamma_{m}$ is the receiving
rate of energy from holographic dark energy to dark matter only.\\

Consider the equation of state:

\begin{equation}
p_{\Lambda}=w_{\Lambda}\rho_{\Lambda},~~p_{m}=w_{m}\rho_{m},~~p_{X}=w_{X}\rho_{X}
\end{equation}

and assume the ratios for energy densities:

\begin{equation}
r_{1}=\frac{\rho_{m}}{\rho_{\Lambda}},~~r_{2}=\frac{\rho_{X}}{\rho_{\Lambda}}
\end{equation}

So from the above continuity equations, we obtain

\begin{equation}
\dot{r}_{1}=r_{1}\Gamma_{\Lambda}+\Gamma_{m}+3H(w_{\Lambda}-w_{m})r_{1}
\end{equation}
and
\begin{equation}
\dot{r}_{2}=(1+r_{2})\Gamma_{\Lambda}-\Gamma_{m}+3H(w_{\Lambda}-w_{X})r_{2}
\end{equation}

Define:

\begin{equation}
w_{m}^{eff}=w_{m}-\frac{\Gamma_{m}}{3r_{1}H}~,~~w_{\Lambda}^{eff}=w_{\Lambda}+\frac{\Gamma_{\Lambda}}{3H}~,
~~w_{X}^{eff}=w_{X}+\frac{\Gamma_{m}-\Gamma_{\Lambda}}{3r_{2}H}
\end{equation}

so that the continuity equations (4) - (6) become
\begin{equation}
\dot{\rho}_{\Lambda}+3H(1+w_{\Lambda}^{eff})\rho_{\Lambda}=0
\end{equation}

\begin{equation}
\dot{\rho}_{m}+3H(1+w_{m}^{eff})\rho_{m}=0
\end{equation}
and
\begin{equation}
\dot{\rho}_{X}+3H(1+w_{X}^{eff})\rho_{X}=0
\end{equation}

Now define the density parameters:

\begin{equation}
\Omega_{m}=\frac{\rho_{m}}{3H^{2}}~,~~\Omega_{\Lambda}=\frac{\rho_{\Lambda}}{3H^{2}}~,
~~\Omega_{X}=\frac{\rho_{X}}{3H^{2}}~,~~\Omega_{k}=\frac{k}{a^{2}H^{2}}
\end{equation}

so from the field equation (2), we obtain

\begin{equation}
\Omega_{m}+\Omega_{\Lambda}+\Omega_{X}=1+\Omega_{k}
\end{equation}

which implies

\begin{equation}
\dot{\Omega}_{m}+\dot{\Omega}_{X}=\dot{\Omega}_{k}-\dot{\Omega}_{\Lambda}
\end{equation}

From equations (9) and (17), we have

\begin{equation}
r_{1}=\frac{\Omega_{m}}{\Omega_{\Lambda}}=\frac{1+\Omega_{k}-\Omega_{\Lambda}-\Omega_{X}
}{\Omega_{\Lambda}}
\end{equation}
and
\begin{equation}
r_{2}=\frac{\Omega_{X}}{\Omega_{\Lambda}}=\frac{1+\Omega_{k}-\Omega_{\Lambda}-\Omega_{m}
}{\Omega_{\Lambda}}
\end{equation}

Now for non-flat universe, the energy density for holographic dark
energy is

\begin{equation}
\rho_{\Lambda}=3c^{2}L^{-2}
\end{equation}

where $c ~(\ge 1)$ is a constant and $L$ represents the radius of
the event horizon measured on the sphere of the horizon defined by

\begin{equation}
L=ar(t)
\end{equation}

where $r(t)$ is a future event horizon obtained from the following
equation

\begin{equation}
r(t)=\frac{\text{sin}~y}{\sqrt{k}}
\end{equation}

where $y=\frac{\sqrt{k}R_{h}}{a}$, $R_{h}$ is the radial size of the event horizon
which is measured in $r$ direction defined by

\begin{equation}
R_{h}=a\int^{\infty}_{t}\frac{dt}{a}
\end{equation}

Now from definition of $\Omega_{\Lambda}$ and using (21), we obtain

\begin{equation}
L=\frac{c}{H\sqrt{\Omega_{\Lambda}}}
\end{equation}

From (21) - (25), we have

\begin{equation}
\dot{L}=\frac{c}{\sqrt{\Omega_{\Lambda}}}-\text{cos}~y
\end{equation}

From (16), (21), (22) and (23), we have

\begin{equation}
\text{cos}~y=\sqrt{1-c^{2}\frac{\Omega_{k}}{\Omega_{\Lambda}}}
\end{equation}

Using (12), (14), (21) and (27), we get the equation of state for holographic dark energy as

\begin{equation}
w_{\Lambda}=-\frac{1}{3}-\frac{2\sqrt{\Omega_{\Lambda}-c^{2}\Omega_{k}}}{3c}-\frac{\Gamma_{\Lambda}}{3H}
\end{equation}

\section{\normalsize\bf{Generalized second law of thermodynamics}}

We consider the FRW universe as a thermodynamical system with the horizon surface
as a boundary of the system. To study the generalized second law (GSL) of thermodynamics
through the universe we deduce the expression for normal entropy using the first law
of thermodynamics i.e., $TdS=PdV+dE$, where, $T,~S,~P,~V$ and $E$ are respectively temperature,
entropy, pressure, volume and internal energy within the event horizon (of radius $L$
which is measured on the sphere of the horizon) of the universe. The entropies for
holographic dark energy, dark matter and other component of dark energy are given by [69]

\begin{equation}
dS_{\Lambda}=\frac{1}{T}(P_{\Lambda}dV+dE_{\Lambda})
\end{equation}

\begin{equation}
dS_{m}=\frac{1}{T}(P_{m}dV+dE_{m})
\end{equation}
and
\begin{equation}
dS_{X}=\frac{1}{T}(P_{X}dV+dE_{X})
\end{equation}

where $V=\frac{4\pi L^{3}}{3}$ is the volume containing matter and dark energies with

\begin{equation}
E_{\Lambda}=\frac{4\pi L^{3}\rho_{\Lambda}}{3}~,~~P_{\Lambda}=w_{\Lambda}^{eff}\rho_{\Lambda}
\end{equation}

\begin{equation}
E_{m}=\frac{4\pi L^{3}\rho_{m}}{3}~,~~P_{m}=w_{m}^{eff}\rho_{m}
\end{equation}
and
\begin{equation}
E_{X}=\frac{4\pi L^{3}\rho_{X}}{3}~,~~P_{X}=w_{X}^{eff}\rho_{X}
\end{equation}

Assuming, $T=\frac{1}{2\pi L}$ and $x=log~a$ and using equations (12), (25), (26), (29) and (32), we obtain

\begin{equation}
\frac{dS_{\Lambda}}{dx}=\frac{24\pi^{2}c^{2}L\dot{L}}{H}\left(w_{\Lambda}+\frac{\Gamma_{\Lambda}}{3H}+\frac{1}{3} \right)
\end{equation}

\begin{equation}
\frac{dS_{m}}{dx}=8\pi^{2}L\left[\left(3w_{m}H-\frac{\Gamma_{m}}{r_{1}} \right)\Omega_{m}L^{2}\dot{L} +
c^{2}\left(\frac{\Omega_{m}\dot{L}}{\Omega_{\Lambda}H}+\frac{L\dot{\Omega}_{m}}{\Omega_{\Lambda}H}
-\frac{L\Omega_{m}}{H\Omega_{\Lambda}^{2}}  \dot{\Omega}_{\Lambda} \right)   \right]
\end{equation}
and
\begin{equation}
\frac{dS_{X}}{dx}=8\pi^{2}L\left[\left(3w_{X}H+\frac{\Gamma_{m}-\Gamma_{\Lambda}}{r_{2}} \right)\Omega_{X}L^{2}\dot{L} +
c^{2}\left(\frac{\Omega_{X}\dot{L}}{\Omega_{\Lambda}H}+\frac{L\dot{\Omega}_{X}}{\Omega_{\Lambda}H}
-\frac{L\Omega_{X}}{H\Omega_{\Lambda}^{2}}  \dot{\Omega}_{\Lambda} \right)   \right]
\end{equation}

Now entropy at the horizon is given by

\begin{equation}
S_{L}=\pi L^{2}
\end{equation}

so that from equations (25) and (26), we obtain

\begin{equation}
\frac{dS_{L}}{dx}=\frac{2\pi c}{H^{2}\sqrt{\Omega_{\Lambda}}}\left(\frac{c}{\sqrt{\Omega_{\Lambda}}}-\text{cos}~y \right)
\end{equation}

From equations (14), (15), (16) and (21), we have

\begin{equation}
3w_{m}H-\frac{\Gamma_{m}}{r_{1}}=-H+\frac{\dot{\Omega}_{\Lambda}}{\Omega_{\Lambda}}-
\frac{\dot{\Omega}_{m}}{\Omega_{m}}-\frac{2H}{c}\sqrt{\Omega_{\Lambda}}~\text{cos}~y
\end{equation}
and
\begin{equation}
3w_{X}H+\frac{\Gamma_{m}-\Gamma_{\Lambda}}{r_{2}}=-H+\frac{\dot{\Omega}_{\Lambda}}{\Omega_{\Lambda}}-
\frac{\dot{\Omega}_{X}}{\Omega_{X}}-\frac{2H}{c}\sqrt{\Omega_{\Lambda}}~\text{cos}~y
\end{equation}

Using equations (16) and (21) and defining the deceleration parameter $q=-1-\frac{\dot{H}}{H^{2}}$ we can obtain

\begin{equation}
\dot{\Omega}_{k}=2qH\Omega_{k}
\end{equation}
and
\begin{equation}
\dot{\Omega}_{\Lambda}=\frac{2\Omega_{\Lambda}}{\Omega_{k}}\left(H\Omega_{\Lambda}-L^{-1}\dot{L}\Omega_{\Lambda}+qH\Omega_{k} \right)
\end{equation}

Using (18), (28), (35)-(41) we get,

\begin{eqnarray*}
\frac{d}{dx}(S_{\Lambda}+S_{m}+S_{X}+S_{L})=\frac{2\pi L\dot{L}}{H}+8\pi^{2}L^{3}\dot{L}\left[ \left( -H+\frac{\dot{\Omega}_{\Lambda}}{\Omega_{\Lambda}}-
\frac{2H}{c}\sqrt{\Omega_{\Lambda}}~\text{cos}~y \right)(\Omega_{m}+\Omega_{X})+(\dot{\Omega}_{\Lambda}-\dot{\Omega}_{k}) \right]
\end{eqnarray*}

\begin{equation}
~~~~~~~~~~~~~~~~~~~~~~~~~~~~~~+8\pi^{2}c^{2}L\left[ \frac{(\Omega_{m}+\Omega_{X})\dot{L}}{\Omega_{\Lambda}H}+\frac{L(\dot{\Omega}_{k}-
\dot{\Omega}_{\Lambda})}{\Omega_{\Lambda}H}
-\frac{L(\Omega_{m}+\Omega_{X})}{H\Omega_{\Lambda}^{2}}  \dot{\Omega}_{\Lambda} \right]
-\frac{16\pi^{2}cL\dot{L}\sqrt{\Omega_{\Lambda}}~\text{cos}~y }{H}
\end{equation}

Now putting the values of $L,~\dot{L},~\text{cos}~y,~\Omega_{X},~\dot{\Omega}_{k}$ and $\dot{\Omega}_{\Lambda}$ from equations (17), (20)-(22), (42) and
(43), we finally get

\begin{eqnarray*}
\frac{d}{dx}(S_{\Lambda}+S_{m}+S_{X}+S_{L})=\frac{2\pi c}{H^{2}\Omega_{k}\Omega_{\Lambda}^{2}}
\left[ -8\pi c(1+\Omega_{k})(\Omega_{\Lambda}^{2}+c^{2}\Omega_{k}^{2})+c\Omega_{k}\Omega_{\Lambda}\{ 1+8\pi(1+c^{2})(1+\Omega_{k}) \}\right.
\end{eqnarray*}
\begin{equation}
\left.
-\Omega_{k}\sqrt{\Omega_{\Lambda}-c^{2}\Omega_{k}}~ \{ \Omega_{\Lambda}+8\pi c^{2}(1+q+\Omega_{k}) \} \right]
\end{equation}

We have seen that r.h.s. of the expression (44) depends on $c,~H,~q,~\Omega_{k}$ and $\Omega_{\Lambda}$. At the present time, setting
$c=1,~\Omega_{k}=0.01$ and $\Omega_{\Lambda}=0.72$, we obtain

\begin{equation}
\frac{d}{dx}(S_{\Lambda}+S_{m}+S_{X}+S_{L})=-\frac{15767+257q}{H^{2}}
\end{equation}

From the above expression we see that $\frac{d}{dx}(S_{\Lambda}+S_{m}+S_{X}+S_{L})\ge 0$ if $q\le -61.43$. But at
the present time, $q>-1$, so GSL can not be satisfied in our model.\\

\section{\normalsize\bf{Discussions}}

In this work, we have considered FRW model of the universe filled with 3 fluids i.e.,
holographic dark energy, dark matter and another unknown component of dark energy.
We have considered a cosmological model of holographic dark energy interacting with
dark matter and another unknown component of dark energy of the universe. We have
assumed two interaction terms $Q$ and $Q'$ in order to include the scenario in which the
mutual interaction between the two principal components (i.e., holographic dark
energy and dark matter) of the universe leads to some loss in other forms of cosmic
constituents. Our model is valid for any sign of $Q$ and $Q'$. If $Q<Q'$, then part
of the dark energy density decays into dark matter and the rest in the other unknown
energy density component. But if $Q>Q'$, then dark matter energy receives from dark
energy and from the unknown component of dark energy. Observation suggests that
dark energy decays into dark matter. We have presented a general prescription of a
cosmological model of dark energy which imposes mutual interaction between holographic
dark energy, dark matter and another fluid. We have obtained the equation of state for
the holographic dark energy density which is interacting with dark matter and other
unknown component of dark energy. Using first law of thermodynamics, we have obtained the
entropies for holographic dark energy, dark matter and other component of dark energy, when
holographic dark energy interacting with two fluids (i.e., dark matter and other component of
dark energy). Also we have found the entropy at the horizon when the radius ($L$) of the event
horizon measured on the sphere of the horizon. We have investigated the GSL of thermodynamics
at the present time for the universe enveloped by this horizon. Finally, it has been obtained
validity of GSL which implies some bounds on deceleration parameter $q$. But at
the present time, $q>-1$, so GSL can not be satisfied in our model.\\\\

{\bf Acknowledgement:}\\

The author is thankful to IUCAA, Pune, India for providing
Associateship Programme under which part of the work was carried
out. The author also thanks to the members of Relativity and
Cosmology Research Centre, Jadavpur University, India for some
illuminating discussions.\\

{\bf References:}\\
\\
$[1]$ N. A. Bachall, J. P. Ostriker, S. Perlmutter and P. J. Steinhardt, \textit{Science} \textbf{284} 1481 (1999).\\
$[2]$ S. J. Perlmutter et al, \textit{Astrophys. J.} \textbf{517} 565 (1999).\\
$[3]$ V. Sahni and A. A. Starobinsky, {\it Int. J. Mod. Phys. A}
{\bf 9} 373 (2000).\\
$[4]$ P. J. E. Peebles and B. Ratra, {\it Rev. Mod. Phys.} {\bf
75} 559 (2003).\\
$[5]$ T. Padmanabhan, {\it Phys. Rept.} {\bf 380} 235 (2003).\\
$[6]$ E. J. Copeland, M. Sami, S. Tsujikawa, {\it Int. J. Mod.
Phys. D} {\bf  15} 1753 (2006).\\
$[7]$ I. Maor and R. Brustein, {\it Phys. Rev. D} {\bf 67} 103508
(2003); V. H. Cardenas and S. D. Campo, {\it Phys. Rev. D} {\bf
69} 083508 (2004); P.G. Ferreira and M. Joyce, {\it Phys.
Rev. D} {\bf 58} 023503 (1998).\\
$[8]$ U. Alam, V. Sahni and A. A. Starobinsky, {\it JCAP} \textbf{0406} 008 (2004).\\
$[9]$ B. Feng, X. L. Wang and X. M. Zhang, \textit{Phys. Lett. B} \textbf{607} 35 (2005).\\
$[10]$ I. Zlatev, L. Wang and P. J. Steinhardt, {\it Phys. Rev.
Lett.} {\bf 82} 896 (1999), {\it astro-ph}/9807002.\\
$[11]$ B. Ratra and P. J. E. Peebles, {\it Phys. Rev. D} {\bf
37} 3406 (1988).\\
$[12]$ T. Chiba, T. Okabe and M. Yamaguchi, {\it Phys. Rev. D}
{\bf 62} 023511 (2000).\\
$[13]$ A. Sen, {\it JHEP} {\bf 0204} 048 (2002).\\
$[14]$ A. Kamenshchik, U. Moschella and V. Pasquier, {\it Phys.
Lett. B} {\bf 511} 265 (2001); V. Gorini, A. Kamenshchik, U.
Moschella and V. Pasquier, gr-qc/0403062; V. Gorini, A. Kamenshchik and U. Moschella, {\it Phys. Rev.
D} {\bf 67} 063509 (2003); U. Alam, V. Sahni, T. D. Saini and A.
A. Starobinsky, {\it Mon. Not. Roy. Astron. Soc.} {\bf 344},
1057 (2003); H. B. Benaoum, {\it hep-th}/0205140; U. Debnath, A.
Banerjee and S. Chakraborty, {\it Class.
Quantum Grav.} {\bf 21} 5609 (2004).\\
$[15]$ R. R. Caldwell, {\it Phys. Lett. B} {\bf 545} 23 (2002).\\
$[16]$ S. M. Caroll, M. Hoffman, M. Trodden, {\it Phys. Rev.
D} {\bf 68} 023509 (2003), {\it astro-ph}/9805201.\\
$[17]$ X. Meng, M. Hu and J. Ren, {\it astro-ph}/0510357.\\
$[18]$ B. McInnes, {\it JHEP} {\bf 0208} 029 (2002).\\
$[19]$ S. Hannestad and E. Mortsell, {\it Phys. Rev. D} {\bf 66} 063508 (2002).\\
$[20]$ A. Melchiori et al, {\it Phys. Rev. D} {\bf 68} 043509 (2003).\\
$[21]$ R. A. Knop et al, {\it Astrophys. J.} {\bf 598} 102 (2003).\\
$[22]$ D. N. Spergel et al, {\it Astrophys. J. Suppl.} {\bf 148}
175 (2003).\\
$[23]$ M. Tegmark et al, {\it Phys. Rev. D} {\bf 69} 103501
(2004).\\
$[24]$ S. Nesseris and L. Perivolaropoulas, {\it Phys. Rev. D}
{\bf 70} 043531 (2004).\\
$[25]$ O. Bertolami, A. A. Sen, S. Sen and P. T. Silva, {\it
Mon. Not. Roy. Astron. Soc.} {\bf 353} 329 (2004).\\
$[26]$ K. Enqvist, S. Hannested and M. S. Sloth, {\it JCAP} {\bf 2} 004 (2005).\\
$[27]$ L. Thorlocius, {\it hep-th}/0404098.\\
$[28]$ G. T. Hooft, {\it gr-qc}/9310026.\\
$[29]$ L. Susskind, {\it J. Math. Phys.} {\bf 36} 6377 (1995).\\
$[30]$ W. Fischler and L. Susskind, {\it hep-th}/9806039.\\
$[31]$ R. Tavakol and G. Ellis, {\it Phys. Lett. B} {\bf 469} 33 (1999).\\
$[32]$ R. Esther and D. Lowe, {\it Phys. Rev. Lett.} {\bf 82} 4967 (1999).\\
$[33]$ E. Verlinde, {\it hep-th}/0008140.\\
$[34]$ B. Wang, Y. Gong and E. Abdalla, {\it Phys. Lett. B} {\bf 624} 141 (2005).\\
$[35]$ Y. Gong, {\it Phys. Rev. D} {\bf 70} 064029 (2004).\\
$[36]$ X. Zhang, {\it Int. J. Mod. Phys. D} {\bf 14} 1597 (2005).\\
$[37]$ D. Pavon and W. Zimdahl, {\it hep-th}/0511053.\\
$[38]$ A. G. Kohen et al., {\it Phys. Rev. Lett.} {\bf 82} 4971
(1999).\\
$[39]$ M. R. Setare, {\it Phys. Lett. B} {\bf 648} 329 (2007).\\
$[40]$ U. Alam, V. Sahni, T. D. Saini and A. A. Starobinsky, {\it Mon. Not. R. Astron. Soc.} {\bf 354} 275 (2004);
D. Huterer and A. Cooray, {\it Phys. Rev. D} {\bf 71} 023506 (2005);
Y. Wang and M. Tegmark, astro-ph/0501351.\\
$[41]$ M. R. Setare and S. Shafei, {\it JCAP} {\bf 09} 011 (2006).\\
$[42]$ M. S. Berger, H. Shojaei, {\it Phys. Rev. D} {\bf 74}
043530 (2006).\\
$[43]$ R. -G. Cai, A. Wang, {\it JCAP}
{\bf 03} 002 (2005).\\
$[44]$ W. Zimdahl, {\it Int. J. Mod. Phys. D} {\bf 14}2319
(2005)\\
$[45]$ B. Hu, Y. Ling, {\it Phys. Rev. D} {\bf 73} 123510
(2006).\\
$[46]$ J. D. Bekenstein, {\it Phys. Rev. D} {\bf 7} 2333 (1973).\\
$[47]$ S. W. Hawking, {\it Commun. Math. Phys.} {\bf 43} 199 (1975).\\
$[48]$ J. M. Bardeen, B. Carter and S. W. Hawking, {\it Commun. Math. Phys.} {\bf 31} 161 (1973).\\
$[49]$ T. Jacobson, {\it Phys. Rev. Lett.} {\bf 75} 1260 (1995).\\
$[50]$ C. Eling, R. Guedens and T. Jacobson, {\it Phys. Rev. Lett.} {\bf 96} 121301 (2006).\\
$[51]$ T. Padmanabhan, {\it Class. Quantum Grav.} {\bf 19} 5387 (2002).\\
$[52]$ R. G. Cai and S. P. Kim, {\it JHEP} {\bf 02} 050 (2005).\\
$[53]$ G. W. Gibbons and S. W. Hawking, {\it Phys. Rev. D} {\bf 15} 2738 (1977).\\
$[54]$ B. Wang, Y. G. Gong and E. Abdalla, {\it Phys. Rev. D} {\bf 74} 083520 (2006).\\
$[55]$ R. Brustein, {\it Phys. Rev. Lett.} {\bf 84} 2072 (2000).\\
$[56]$ M. Li, {\it Phys. Lett. B} {\bf 603} 1 (2004).\\
$[57]$ P. F. Gonzalez-Diaz, hep-th/0411070.\\
$[58]$ P. C. W. Davies, {\it Class. Quantum Grav.} {\bf 4} L225 (1987).\\
$[59]$ P. C. W. Davies, {\it Class. Quantum Grav.} {\bf 5} 1349 (1988).\\
$[60]$ R. Bousso, {\it Phys. Rev. D} {\bf 71} 064024 (2005).\\
$[61]$ G. Izquierdo and D. Pavon, astro-ph/0505601.\\
$[62]$ G. Izquierdo and D. Pavon, {\it Phys. Rev. D} {\bf 70} 127505 (2004).\\
$[63]$ S. Nojiri and S. D. Odinstov, hep-th/0506212.\\
$[64]$ Q. Huang and M. Li, {\it JCAP} {\bf 0408} 013 (2004).\\
$[65]$ I. Brevik, S. Nojiri, S.D. Odintsov and L. Vanzo, {\it Phys. Rev. D} {\bf 70}
043520 (2004).\\
$[66]$ P. F. Gonzalez-Diaz and C. L. Siguenza, {\it Nucl. Phys. B} {\bf 697} 363 (2004); E.
Babichev, V. Dokuchaev and Y. Eroshenko, {\it Phys. Rev. Lett.} {\bf 93} 021102
(2004).\\
$[67]$ Y. Gong, B. Wang and A. Wang, {\it JCAP} {\bf 01} 024 (2007).\\
$[68]$ N. Cruz, S. Lepe and F. Pe$\tilde{\text{n}}$a, {\it Phys. Lett. B} {\bf 663} 338 (2008); arXiv:0804.3777 [hep-th].\\
$[69]$ M. R. Setare, {\it JCAP} {\bf 01} 023 (2007); M. R. Setare and S. Shafei, {\it JCAP} {\bf 09} 011 (2006).\\

\end{document}